\documentclass[a4paper, conference]{IEEEtran}

\usepackage[utf8]{inputenc}
\usepackage[english]{babel}

\usepackage{pifont}
\usepackage[nomain,nopostdot,acronym]{glossaries}
\usepackage{subcaption,graphicx}
\usepackage{graphicx}
\usepackage{subcaption}
\usepackage{textcomp}
\usepackage{xcolor}
\IEEEoverridecommandlockouts
\usepackage{graphicx}
\usepackage{dblfloatfix}

\usepackage[final]{pdfpages}
\usepackage[most]{tcolorbox}
\usepackage{xargs}
\usepackage{wrapfig}
\usepackage[export]{adjustbox}
\usepackage{tikz}
    \usetikzlibrary{shapes.arrows}

\usepackage{booktabs}
\usepackage{threeparttable}

\usepackage{pgfplots}
\usepackage{adjustbox}
\pgfplotsset{
    compat=1.16,
    tick label style={font=\tiny},
    label style={font=\tiny},
}
\usepackage{pgfplots}
\usepackage{adjustbox}
\usepackage[colorinlistoftodos,prependcaption,textsize=tiny]{todonotes}
\usepackage{glossaries}

\usepackage[left=1.22cm,right=1.22cm,top= 2.24cm, bottom=4.85cm]{geometry}

\def \tfwidth{0.99\linewidth}
\def \tfheight {0.6\linewidth}

\newcommandx{\unsure}[2][1=]{\todo[linecolor=red,backgroundcolor=red!25,bordercolor=red,#1]{#2}}
\newcommandx{\change}[2][1=]{\todo[linecolor=blue,backgroundcolor=blue!25,bordercolor=blue,#1]{#2}}
\newcommandx{\info}[2][1=]{\todo[linecolor=lime,backgroundcolor=lime!25,bordercolor=lime,#1]{#2}}
\newcommandx{\improvement}[2][1=]{\todo[linecolor=Plum,backgroundcolor=Plum!25,bordercolor=Plum,#1]{#2}}
\newcommandx{\thiswillnotshow}[2][1=]{\todo[disable,#1]{#2}}

\begin{document}
\setlength{\abovecaptionskip}{10pt plus 3pt minus 2pt}

\title{Near Real-Time Data-Driven Control of Virtual Reality Traffic in Open Radio Access Network
\thanks{Andreas Casparsen (aca@es.aau.dk), Beatriz Soret (bsa@es.aau.dk), Jimmy Jessen Nielsen, and Petar Popovski (petarp@es.aau.dk) are with the Department of Electronic Systems, Aalborg University, Denmark. Beatriz Soret is also with the Department of Communications Engineering, Universidad de Málaga, Spain. This work was partly funded by the European Commission as part of the IntellIoT project, under the H2020 framework grant no. 957218.}}

\author{Andreas Casparsen, Beatriz Soret, Jimmy Jessen Nielsen, and Petar Popovski}


\maketitle
\thispagestyle{plain}
\pagestyle{plain}
\newacronym{ric}{RIC}{RAN Intelligent Controller}
\newacronym{ran}{RAN}{Radio Access Network}
\newacronym{ue}{UE}{User Equipment}
\newacronym{mno}{MNO}{Mobile Network Operator}
\newacronym{cots}{COTS}{Common Off The Shelf}
\newacronym{qoe}{QoE}{Quality of Experience}
\newacronym{qos}{QoS}{Quality of Service}
\newacronym{vr}{VR}{Virtual Reality}
\newacronym{oai}{OaI}{OpenAirInterface}
\newacronym{mcs}{MCS}{Modulation and Coding Scheme}
\newacronym{dlt}{DLT}{Distributed Ledger Technology}
\newacronym{mec}{MEC}{Multi-Access Edge Computing}
\newacronym{sdr}{SDR}{Software defined radio}
\newacronym{cqi}{CQI}{Channel Quality Indicator}
\newacronym{urllc}{URLLC}{Ultra Reliable Low Latency Communication}
\newacronym{embb}{eMBB}{enhanced Mobile Broadband}
\newacronym{mmtc}{mMTC}{Massive Machine-Type Communications}

\newacronym{rb}{RB}{Resource Block}

\newacronym{powder}{POWDER}{ Platform for Open Wireless Data-driven Experimental Research}
\newacronym{sm}{SM}{Service Model}
\newacronym{kpi}{KPI}{key performance indicator}
\newacronym{capex}{CapEx}{Capital Expenditure}
\newacronym{opex}{OpEx}{Operational Expenditure}
\newacronym{fps}{FPS}{Frames Per Second}
\newacronym{cbr}{CBR}{Constant Bit Rate}
\newacronym{mac}{MAC}{Medium Access Control}
\newacronym{KQI}{KQI}{Key Quality Indicators}
\newacronym{RSRP}{RSRP}{Reference Signal Received Power}
\newacronym{RSRQ}{RSRQ}{Reference Signal Received Quality}
\newacronym{TTI}{TTI}{Transmission Time Interval}
\newacronym{AI}{AI}{Artificial Intelligence}

\newacronym{cu}{CU}{Central Unit}
\newacronym{du}{DU}{Distributed Unit}
\newacronym{ocu}{O-CU}{O-RAN Central Unit}
\newacronym{odu}{O-DU}{O-RAN Distributed Unit}
\newacronym{sdran}{SD-RAN}{Software-Defined RAN}
\newacronym{rbg}{RBG}{Resource Block Group}

\newacronym{ru}{RU}{Radio Unit}
\newacronym{oran}{ORAN}{Open Radio Access Network}

\newacronym{RMS}{RMS}{Root Mean Square}
\newacronym{UDP}{UDP}{User Datagram Protocol}
\newacronym{UPF}{UPF}{User Plane Function}
\newacronym{TCP}{TCP}{Transmission Control Protocol}

\noindent
\begin{abstract}

In mobile networks, \gls{oran} provides a framework for implementing network slicing that interacts with the resources at the lower layers. Both 
monitoring and \gls{ran} control is feasible for both 4G and 5G systems. In this work, we consider how data-driven resource allocation in a 4G context can enable adaptive slice allocation to steer the experienced latency of \gls{vr} traffic towards a requested latency. We develop an xApp for the near real-time \gls{ric} that embeds a heuristic algorithm for latency control, aiming to: (1) maintain latency of a VR stream around a requested value; and (2) improve the available \gls{ran} allocation to offer higher bit rate to another user.
We have experimentally demonstrated the proposed approach in an \gls{oran} testbed. Our results show that the data-driven approach can dynamically follow the variation of the traffic load while satisfying the required latency. This results in 15.8\% more resources to secondary users than a latency-equivalent static allocation.

\glsresetall
\end{abstract}

\renewcommand\IEEEkeywordsname{Keywords}
\noindent
\begin{IEEEkeywords}
\glsresetall
VR, ORAN, Network slicing, xApp
\end{IEEEkeywords}

\ifCLASSOPTIONpeerreview
\begin{center} \bfseries EDICS Category: 3-BBND \end{center}
\fi
%
\IEEEpeerreviewmaketitle

\section{Introduction}

\begin{figure}[t!]
    \centering
    \includegraphics[width=\linewidth,keepaspectratio]{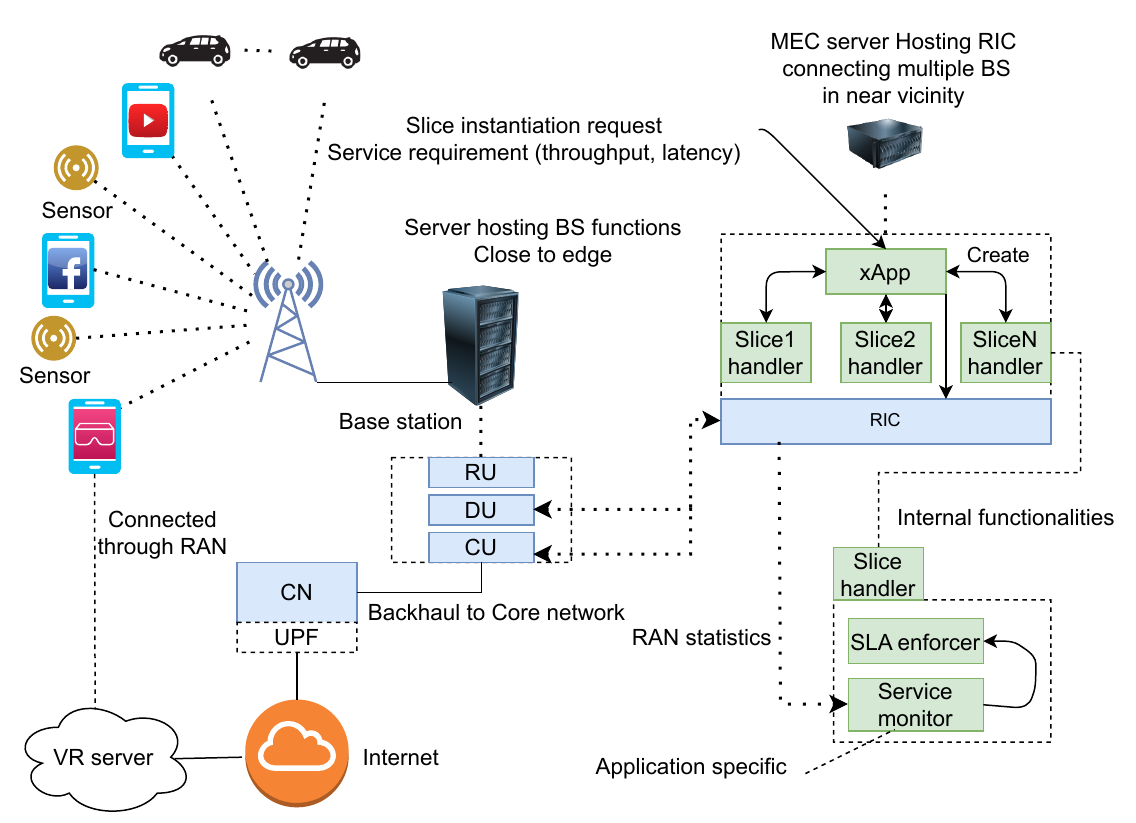}
    \caption{System model}
    \vspace{-2.5em}
    \label{fig:systemmodel}
\end{figure}

While the 5G networks are being rolled out as the new cellular generation, a new paradigm is being developed for how the \gls{ran} should be defined architecturally.
Unlike the classical closed \gls{ran} approach where the \gls{mno} buys an all-in-one solution from a vendor. Instead, \gls{oran} provides an open architecture based on disaggregation, intelligent control, virtualization, and open interfaces. The new open design is intended to enable interoperability among vendors, therefore the economic and technological potential is huge. Indeed, \gls{oran} is of great interest to \glspl{mno}, who find the status quo closed \gls{ran} to be expensive in terms of \gls{capex} and \gls{opex}. They would always have to buy the full implementation of a base station from a single vendor \cite{azariah2022survey}. 
In this regard, Rakuten has demonstrated a reduced cost of operation \cite{9764985}. \gls{capex} was reduced by $40$\% while site equipment and deployment costs were reduced by $60$\% and $50$\%, respectively, compared to traditional \gls{ran}.
To ensure compatibility with 3GPP systems, \gls{oran} only adds a new set of interfaces, independent of the 3GPP specifications connecting the different \gls{ran} components.
No changes are required to 3GPP specifications, rather, interoperability between equipment from multiple vendors is desired through new open interfaces. 
A key aspect of \gls{oran} is to integrate intelligence into the \gls{ran}. This is the role of the \gls{ric}, a software-defined component responsible for controlling and optimizing the \gls{ran}. The \gls{ric} enables control loops on two levels: near real-time (near-RT), operating on a ten-millisecond to one-second level, and non real-time (non-RT), with no timing requirements. The monitoring and control are done via software applications called xApps (for near-RT) and rApps (for non-RT).
The near-RT \gls{ric} utilizes the E2 interface to communicate with E2 nodes i.e. eNB in 4G and  \gls{odu} and \gls{ocu} in 5G. 
The E2 interface is used by xApps to monitor and execute fast control of the \gls{ran} on a smaller set of base stations, e.g., for scheduling, RAN slicing, load balancing, and handover.
The non-RT \gls{ric} interacts with a greater amount of base stations and uses rApps for non-RT control of eNBs, \gls{odu}, and \gls{ocu} through the O1 interface. Typical use cases are for long-term learning and monitoring for policy-making on a networking level as well as assisted machine learning management. Policies from rApps may in this sense also influence how xApps make their decisions. Further, the non-RT \gls{ric} can communicate directly with the near-RT \gls{ric} via the A1 interface \cite{polese2023understanding}.
The ORAN framework enriches the potential of typical 5G \gls{ran} slicing towards supporting heterogeneous applications. Mixed requirements for low latency, high bandwidth and massive access as illustrated in Fig. \ref{fig:systemmodel} can be enhanced by RAN intelligence. While the canonical services of 5G: \gls{urllc}, \gls{mmtc} and \gls{embb} have straightforward requirements, it can be more difficult to know how the \gls{ric} can optimize the user experience of a specific application or service. An example is \gls{vr}-traffic as it has a mixture of low latency and high bandwidth requirements in addition to a varying data rate despite the use of \gls{cbr} video coding \cite{chiariotti2022temporal}. Obviously, a dynamic real-time adaptation of resources is necessary to fully satisfy such application requirements. 

In the context of \gls{ran}-slicing, 5G is a typical association. However, \gls{ran}-slicing is not necessarily a 5G specific functionality. Paradigms for software-controlling the \gls{ran} exist such as the FlexRAN controller \cite{flexran} for a \gls{sdran} with 4G. The principle is that the \gls{ran} can be controlled and updated in terms of its functionalities and resource allocation by isolating resources for specific groups of users.
The \gls{sdran} principle is leveraged in the \gls{oran} paradigm, where xApps are utilized for resource control for both 4G and 5G.
In \cite{orhan2021connection}, a control loop adapts \gls{ue} association to base stations aiming to optimize for sum throughput, base station coverage, and even load distribution. This is feasible due to utilizing \gls{ran}-level information on link quality and evaluating instantaneous base station load.
The modularity of \gls{oran} is shown in 
\cite{dryjanski2021toward} through multiple xApps striving to achieve a singular goal of traffic steering. In the work three xApps were created, one for spectrum management, cell UE association, and resource allocation. These act independently with different actions to optimize individual goals. They could be altered by the authors, as different policies may change over time.
A predictive uplink slicing xApp is proposed in \cite{wiebusch2023towards}, where experimental evaluation with 4G components is used. The work considers \gls{embb} and \gls{urllc} traffic for two users, where the xApp seeks to guarantee the latency of the \gls{urllc} traffic, while allocating as many \glspl{rb} as possible to the best-effort traffic in the \gls{embb} slice. 
In \cite{polese2022colo} the Colosseum platform is used to emulate a \gls{ran}, for which specialized schedulers are trained (both online and offline) to optimize metrics associated with three different slices, \gls{embb}, \gls{urllc}, and \gls{mmtc}.
The work illustrates how specialization of logical functions can improve the service for the \glspl{ue}, and how xApps can help realise this.
In \cite{johnson2022nexran} the POWDER testbed, based on the srsRAN project, is used for 4G RAN slicing. The authors have applied RAN control through a RESTful interface to make slice requests. The authors implement an xApp that applies slicing policies. These adapts resource allocation according to user throughput to balance sub-frame allocation.  
In \cite{nguyen2023network} optimization of flow splitting, congestion control and scheduling is proposed for traffic steering in OpenRAN. Historical data is used to split data to the \glspl{ue} on the non-RT RIC. Meanwhile, congestion control of queues is processed on the near-RT \gls{ric}.
In \cite{kavehmadavani2023intelligent} traffic prediction is used along with user association and radio resource management for the process of traffic steering of \gls{embb} and \gls{urllc} traffic. \gls{AI} is used for the traffic prediction in the non-RT \gls{ric}. Given the traffic prediction, \glspl{rb} are allocated to \glspl{ue}. Latency \glspl{qos} guarantees are given through a latency threshold for both \gls{embb} and \gls{urllc} traffic.
While these traffic steering approaches do improve the experience of different types of traffic, they do so only from a resource allocation perspective and not from an application perspective.


A different perspective is provided in \cite{baena2020estimation}, where the authors seek to predict video \gls{KQI} of a video feed for different network configuration choices and configure network slices accordingly. The ML-based prediction uses both video \glspl{KQI} as defined by 3GPP in addition to \gls{RSRP} and \gls{RSRQ} radio-level information.
In \cite{baena2022measuring}, a similar approach is taken to estimate visual and interaction \glspl{KQI} in Cloud gaming scenarios and in turn optimize resource allocation for \gls{qoe}. First, A statistical analysis of application and radio-level \glspl{KQI} is used to rank the features of importance. Second, six different ML-based methods are compared in terms of prediction accuracy, considering the cases of measurements being taken on the UE or BS side.
These works combine high-level service metrics with additional RAN-level metrics to estimate the application performance. These do, however, not address \gls{ran} control to approach a desired performance. Neither do they utilize the \gls{oran} framework for a \gls{sdran} with rich monitoring.

While the above works concerned with network slicing and traffic steering have successfully demonstrated per-slice optimization of performance metrics such as latency and throughput in \gls{oran}, the characteristics of the end user application is not considered in any of the optimizations, when applying RAN-slicing.
On the other hand, existing work has also shown how \gls{qoe} estimation also can leverage radio information for end-to-end performance estimation. For this, the knowledge of the application is of importance for the \gls{kpi} measurement. This has, however, not been done in an adaptive way, nor in an \gls{oran} context.
The openness of the \gls{oran} framework suggests that different tasks can be outsourced to specialists. 
That is, different vendors can provide specific xApps for serving a variety of heterogeneous services to be supported by the \gls{ran} as seen in Fig. \ref{fig:systemmodel}. These could provide application-layer \gls{qoe} guarantees through \gls{ran} control for specific applications that have complex requirements.
In \cite{giupponi2022blockchain} a framework for \gls{dlt} management of RAN sharing is proposed. Service providers facilitate their service on top of the general \gls{ran} infrastructure, by renting resources.
In this context, radio resource management has a different use case than otherwise. While network slicing for \gls{embb} or \gls{urllc} service require the optimization of basic metrics such as throughput or latency, respectively. Such simple optimization would not necessarily lead to good performance for applications with more complex needs.
In the present work, we propose a data-driven, application-aware approach to optimizing the experienced latency of a \gls{vr} application, while limiting overprovisioning. Our proposed algorithm is implemented as an xApp and evaluated experimentally in our ORAN testbed. 

\section{System model}
The considered system consists of $N$ \glspl{ue} with heterogeneous services and requirements connected to a base station. The base station possesses a near-RT \gls{ric} which exposes control and monitoring functionalities.
The xApp is the key component of our design. When a service is requested, the xApp requests a service slice and instantiates a slice handler, which has the task of monitoring and estimating the \gls{ue} application performance. The estimation reports are used to allocate more or fewer resources for the guarantees that were made for the corresponding slice. The request for a slice creation is from an external entity to the \gls{ran} such as a service provider. Fig. \ref{fig:systemmodel} illustrates this association and interaction.
The handlers individually monitor and estimate the performance of the slice that they control. They do so by observing the \gls{ran}-side statistics on data sent to the specific \gls{ue}. They report if more or fewer resources are required for the service. If resources are insufficient, the xApp will handle conflict mitigation and interact with the \gls{ric} for resource allocation.
Ultimately, the xApp will attempt to uphold service guarantees, but when this is infeasible a decision must be made to prioritize the different slices.
We consider that our slice handler uses an \gls{sm} to continuously acquire fresh \gls{ran} data, and is specialized for inference of the service in question.
Thus, different modules could be developed by different service providers. Both to infer on the application performance, as well as estimating the required resources to uphold a guarantee.
In this work, we consider an xApp that makes inference on the characteristics of \gls{vr} traffic.
With this, slice guarantees for a service is made feasible through resource isolation in the \gls{ran}
In this regard, a request on a video-frame latency on average is requested, while background traffic occurs. 
Optimally the latency guarantee is close to the latency request, while as much bandwidth is allocated to another user as possible.
We consider this as background data that could be available to other users, hence maximizing this means being more effective with the available bandwidth resources.


\section{Algorithmic solution} \label{sec:solution}
In this section, we define the principles of the data-driven process for inferring the \gls{vr} frame latency, and how these are used to steer the allocation.
For the data-driven allocation to ensure the latency experienced by a \gls{vr} UE, we consider an algorithm that uses \gls{ran} data. 
The data provided for the algorithm is the downlink bits sent to the \gls{ue} from the base station per millisecond. 
We define a heuristic algorithm that detects individual video frames sent to the \gls{ue} from the spacing of the video \gls{fps}.
Once every time interval, the corresponding aggregate \gls{mac} data samples of how many bits were sent to the UE and when, is fed to the algorithm. Every transmission of bits following one or more empty slots is marked as a chunk of 1 or more video frames. In the second step, the algorithm considers the expected number of video frames, given the used video \gls{fps}, by which consecutive, partially overlapping video frames are detected. 
With the set of frames detected over the past one-second period, we calculate an \textbf{average\_latency}. This is computed as the time it takes to send the frame, and adding an empirically observed offset to the latency based on the network.
This offset is a compound average value describing propagation delay, networking delay and processing delay and other inherent delays in the system. It was found as the average offset from using the algorithm to estimate latency of the video frames, and the observed latency from server to client.
The latency is averaged over the number of frames observed in the period. The result is used to decide if the allocation should be modified.
For this purpose, we add an allowed slack to the guarantee. That is, if $average\_latency > requested\_latency + slack$ is true, we increase the \gls{rb} allocation. Oppositely, if $average\_latency < requested\_latency - slack$ is true we decrease the \gls{rb} allocation. Otherwise, the resource allocation remains unchanged.


\section{Experimental setup and Implementation}
\begin{figure}[t]
    \centering
    \includegraphics[height=9cm,keepaspectratio]{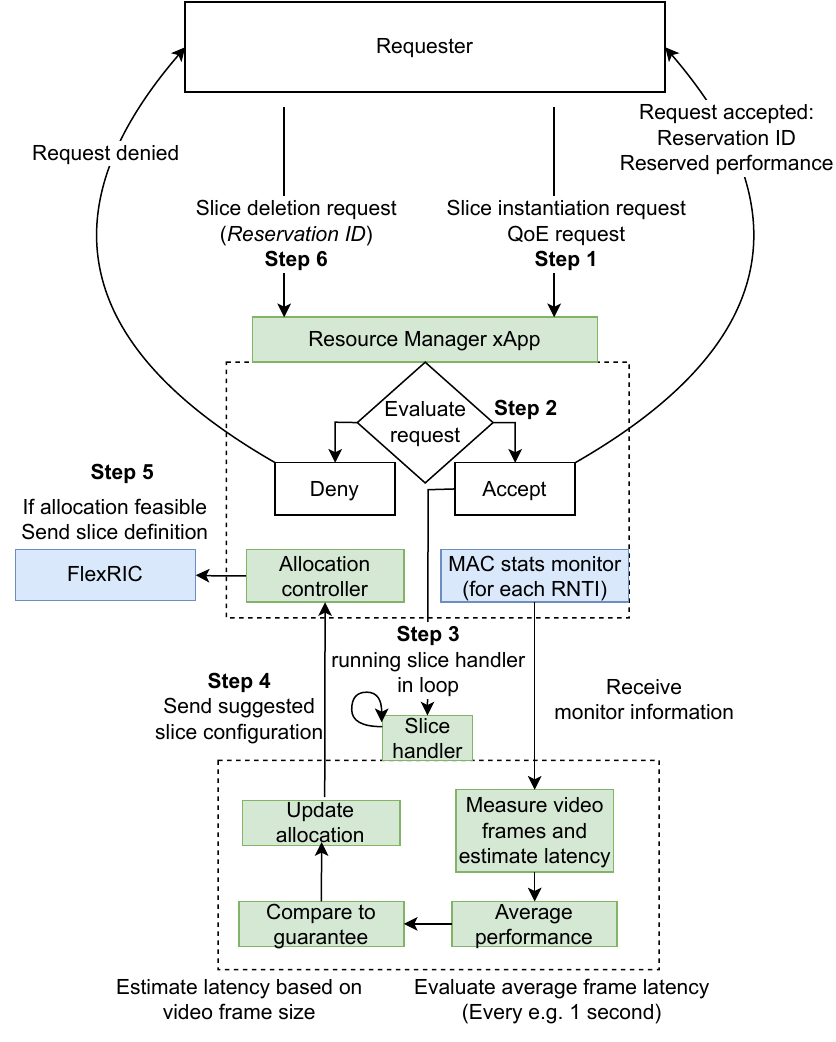}
    \caption{System logic and flow of the xApp.}
    \vspace{-2em}
    \label{fig:slice_creation}
\end{figure}

In this section, we define the xApp to be implemented with basic functionalities to control the \gls{ran} through slicing. This is executed via a Python application to create an interface to make requests.
Upon the request of a slice with a latency requirement, a handler is created and attached to the specific slice, on which the \gls{ue} is associated. The handler implements the algorithmic solution in section \ref{sec:solution}. 
How well latency fulfilment can be supported will depend on the data-driven approach. As the required bandwidth varies over time, it may not always be feasible, if more users require bandwidth.
In 5G low-latency can be supported in an \gls{urllc} slice with short \gls{TTI}. As we base our implementation on \gls{oai} \cite{oai}, and use the FlexRIC implementation \cite{flexric} from \cite{schmidt2021flexric}, only 4G slicing is currently possible.
With our experimental setup based on 4G with \gls{oran} support, we optimize the use of resources by exploiting the control options of \gls{oran} to optimize the resource allocation. Since network slicing is a service provided for both 4G and 5G, the learnings from this work for 4G can also be applied to 5G.
We employ the algorithmic solution described in section \ref{sec:solution}. The xApp will receive a request for a latency guarantee. Its control loop will then continuously allocate \glspl{rb} to keep the VR latency around the request.
On the MAC layer, information relating to the signal channel quality, and data sent to and from the \gls{ue} can be extracted using the default monitoring \gls{sm} implemented in the FlexRIC project. When instantiating a \gls{sm}, different update rates can be used for how often function calls are made. For our purpose, we set our monitoring to receive \gls{ran} statistics every one millisecond.
The flow of the system is illustrated on Fig. \ref{fig:slice_creation}. It illustrates the process through which a slice is created, and how the handler ensures a specific latency in this data-driven fashion.


In \textbf{step one}, a message is sent to create a slice with the bit rate of the video, the \gls{fps}, and the desired latency. \textbf{Step two}, the slice is evaluated to be accepted, or denied, based on if the bit rate can be supported. \textbf{Step three} occurs if the request is accepted, which leads to creating the slice and the handler.
The handler will execute in a loop until the removal of the slice. Measurements of the RAN traffic, inference on video frames, and estimates of the latency are performed before an averaging is applied to the estimated video latencies. Depending on the result, the allocation may be updated if too much or too little was allocated to fulfil the latency.
\textbf{Step four} occurs if an allocation update was found necessary, which means the xApp sends a request to update the allocation made. 
\textbf{Step five} occurs at the allocation controller as information from handlers is gathered, which is used to define to whom to allocate more or fewer resources. It is determined if the current bandwidth allocation is sufficient or if it should be altered. New slice allocation schemes are then sent to the FlexRIC.
\textbf{Step 3} continues its loop, and \textbf{steps 4 and 5} will continue to update resource allocation until \textbf{step 6}, where a request to delete the slice is received. The request contains a slice ID, which acts as an identifier for the service request.

The solution is tested experimentally by deploying open-source components. The implementation by \gls{oai} \cite{oai} is used for the base station, srsRAN \cite{srsran} for the \glspl{ue}, and Open5GS \cite{open5gs} for a non-standalone core network.
The \gls{ric} is deployed as the FlexRIC \cite{flexric}, which currently supports 4G slicing functionalities for both \gls{oai} and srsRAN.
Each of the two \glspl{ue} utilizes a USRP B210 as radio front-end. One is used as for the \gls{vr} user, the other for the secondary \gls{ue} where we run iperf3 to generate background traffic.
We deploy base station, core network, FlexRIC, and xApp on the same physical computer in a dockerized environment. 
The eNB uses a USRP X310 radio front-end that operates with 20 MHz of bandwidth. This translates into 100 \glspl{rb} (that can be allocated in \glspl{rbg} of 4 RBs) to be allocated in the downlink.
The UE USRPs are connected to the eNB USRP using cable-connectors. 
The resulting channels are randomly time-varying.
For the considered configuration the average downlink data rate from the eNB, shared between the two users, is $\sim 34$Mbit/s.
The used \gls{vr} traffic trace files \cite{chiariotti2022temporal} contain the size of the video frame and the timestamp that they were generated.
We have developed a trace playback server that is instantiated on the \gls{UPF} on the core network. It transmits the frames at their respective timestamps in \gls{UDP} packets to the user, who has a client running.
The client can determine the latency that it took the frame to traverse the network using the timestamp encoded in the packet the client receives.
For accurate latency measurements, the two computers are clock-synchronized with \textbf{chrony}. The \gls{RMS} clock offset of the system from the \gls{ue} computer, to the other is less than 50 $\mu$s, which is sufficient for measuring millisecond-level latency.
In addition to the VR traffic traces sent to one UE, a full-buffer traffic flow is generated using the \textbf{iperf3} traffic generator. This setup allows to evaluate the latency of \gls{vr} frames in a fully loaded system, while measuring leftover throughput in the network for secondary users. 
 
The assessment of the data-driven approach is twofold. First, we look at the latency. For VR latency we estimate the video frames latency at the xApp based on the \gls{ran} data and a one second moving average as explained in section \ref{sec:solution}, assuming a slack of 1 millisecond. This is compared with the latency measured at the UE to assess how accurate the estimation is. Second, we evaluate the bit rate of the secondary \gls{ue}, which receives iperf3 traffic from the eNB to quantify the efficiency of slicing strategy.

\begin{figure*}[b!]
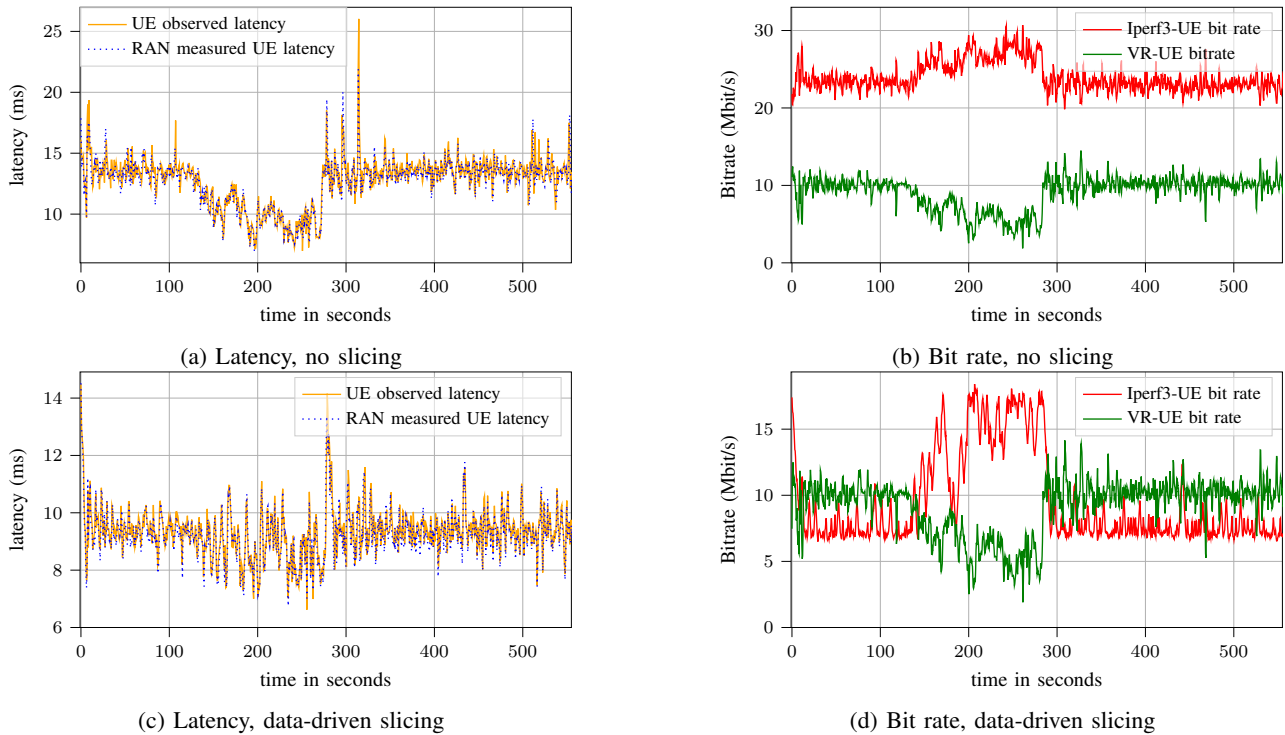

        \begin{subfigure}[b]{.5\linewidth}
	    \centering
        \input{Figures/tikz/resultnaRBs_vr_plotter}
        \caption{Latency, no slicing}
        \label{fig:vr_ue_1}
    \end{subfigure}
	\begin{subfigure}[b]{.5\linewidth}
	    \centering
        \input{Figures/tikz/resultnaRBs_data_plotter}
        \caption{Bit rate, no slicing}
        \label{fig:data_ue_1}
    \end{subfigure}
        \begin{subfigure}[b]{.5\linewidth}
	    \centering
        \input{Figures/tikz/result10ms_vr_plotter}
        \caption{Latency, data-driven slicing}
        \label{fig:vr_ue_3}
    \end{subfigure}
	\begin{subfigure}[b]{.5\linewidth}
	    \centering
        \input{Figures/tikz/result10ms_data_plotter}
        \caption{Bit rate, data-driven slicing}
        \label{fig:data_ue_3}
    \end{subfigure}

    \vspace{-0.5em}
     \caption{Performance of the two users for the no slicing and data-driven slicing cases.}
     
\end{figure*}

\section{Results and Discussion} \label{sec:results}

For the experiments, the \gls{vr} server streams a 60 \gls{fps} VR video with a 10 Mbit/s avg. bit rate in the downlink to a client UE, while the iperf3 traffic generator has a full buffer traffic model, meaning that the traffic streams must share the available downlink bandwidth. 
Initially, we consider a shared resource approach, with no slicing. From this, we extract the latency of the video frames for the \gls{vr} traffic, and the downlink data rate of the other \gls{ue}, which are the key performance indicators of interest.
In Fig. \ref{fig:vr_ue_1} an overview of latency over time can be seen. 
First, we study the accuracy of the of UE video frame latency estimation. Fig. \ref{fig:vr_ue_1} shows that the estimated \gls{ran} latency follows the actually observed latency from the \gls{ue} well, thereby confirming the ability to estimate the latency well.
From around the 280-second mark, the latency increases to around 13-14 ms, due to an increase in VR-UE bit rate as shown in Fig. \ref{fig:data_ue_1} (a) and (b). Obviously, the impact on the iperf3 \gls{ue} is a decrease in the traffic rate it can transmit. 

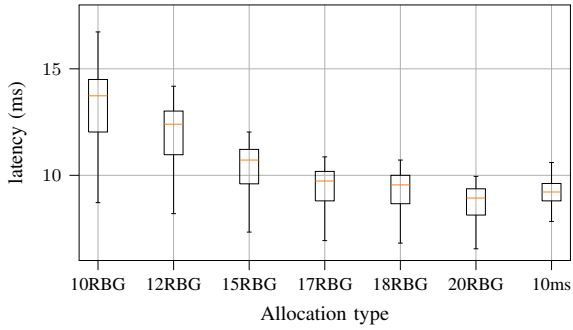
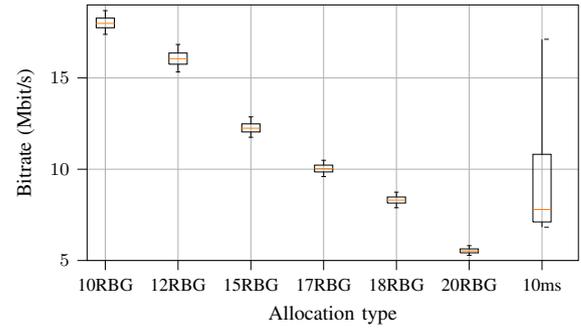
\begin{figure*}
        \begin{subfigure}[b]{.5\linewidth}
	    \centering
\begin{tikzpicture}[scale=0.85]

\definecolor{darkgray176}{RGB}{176,176,176}
\definecolor{darkorange25512714}{RGB}{255,127,14}


\begin{axis}[
width=\tfwidth,
height=\tfheight,
legend cell align={left},
legend style={fill opacity=0.6,font={\tiny} ,draw opacity=1, text opacity=1, draw=lightgray204},
tick align=outside,
tick pos=left,
x grid style={darkgray176},
xlabel={Allocation type},
xmin=-2, xmax=50,
xtick style={color=black},
xtick={0,8,16,24,32,40,48},
xticklabels={10RBG, 12RBG, 15RBG, 17RBG, 18RBG, 20RBG, 10ms},
y grid style={darkgray176},
ylabel={latency (ms)},
ymin=6, ymax=18,
xmajorgrids,
ymajorgrids,
ytick style={color=black},
label style={font=\small},
tick label style={font=\footnotesize}
]

\addplot [black]
table {%
-1 12.0333333333333
1 12.0333333333333
1 14.5
-1 14.5
-1 12.0333333333333
};
\addplot [black]
table {%
0 12.0333333333333
0 8.71666666666667
};
\addplot [black]
table {%
0 14.5
0 16.7333333333333
};
\addplot [black]
table {%
-0.25 8.71666666666667
0.25 8.71666666666667
};
\addplot [black]
table {%
-0.25 16.7333333333333
0.25 16.7333333333333
};
\addplot [black]
table {%
7 10.9666666666667
9 10.9666666666667
9 13.0166666666667
7 13.0166666666667
7 10.9666666666667
};
\addplot [black]
table {%
8 10.9666666666667
8 8.2
};
\addplot [black]
table {%
8 13.0166666666667
8 14.1833333333333
};
\addplot [black]
table {%
7.75 8.2
8.25 8.2
};
\addplot [black]
table {%
7.75 14.1833333333333
8.25 14.1833333333333
};
\addplot [black]
table {%
15 9.6
17 9.6
17 11.2166666666667
15 11.2166666666667
15 9.6
};
\addplot [black]
table {%
16 9.6
16 7.33333333333333
};
\addplot [black]
table {%
16 11.2166666666667
16 12.0333333333333
};
\addplot [black]
table {%
15.75 7.33333333333333
16.25 7.33333333333333
};
\addplot [black]
table {%
15.75 12.0333333333333
16.25 12.0333333333333
};
\addplot [black]
table {%
23 8.8
25 8.8
25 10.1833333333333
23 10.1833333333333
23 8.8
};
\addplot [black]
table {%
24 8.8
24 6.93333333333333
};
\addplot [black]
table {%
24 10.1833333333333
24 10.8666666666667
};
\addplot [black]
table {%
23.75 6.93333333333333
24.25 6.93333333333333
};
\addplot [black]
table {%
23.75 10.8666666666667
24.25 10.8666666666667
};
\addplot [black]
table {%
31 8.66666666666667
33 8.66666666666667
33 10
31 10
31 8.66666666666667
};
\addplot [black]
table {%
32 8.66666666666667
32 6.81666666666667
};
\addplot [black]
table {%
32 10
32 10.7166666666667
};
\addplot [black]
table {%
31.75 6.81666666666667
32.25 6.81666666666667
};
\addplot [black]
table {%
31.75 10.7166666666667
32.25 10.7166666666667
};
\addplot [black]
table {%
39 8.13333333333333
41 8.13333333333333
41 9.36666666666667
39 9.36666666666667
39 8.13333333333333
};
\addplot [black]
table {%
40 8.13333333333333
40 6.55
};
\addplot [black]
table {%
40 9.36666666666667
40 9.95
};
\addplot [black]
table {%
39.75 6.55
40.25 6.55
};
\addplot [black]
table {%
39.75 9.95
40.25 9.95
};
\addplot [black]
table {%
47 8.8
49 8.8
49 9.61666666666667
47 9.61666666666667
47 8.8
};
\addplot [black]
table {%
48 8.8
48 7.83333333333333
};
\addplot [black]
table {%
48 9.61666666666667
48 10.6
};
\addplot [black]
table {%
47.75 7.83333333333333
48.25 7.83333333333333
};
\addplot [black]
table {%
47.75 10.6
48.25 10.6
};
\addplot [darkorange25512714]
table {%
-1 13.7333333333333
1 13.7333333333333
};
\addplot [darkorange25512714]
table {%
7 12.4
9 12.4
};
\addplot [darkorange25512714]
table {%
15 10.7166666666667
17 10.7166666666667
};
\addplot [darkorange25512714]
table {%
23 9.73333333333334
25 9.73333333333334
};
\addplot [darkorange25512714]
table {%
31 9.55
33 9.55
};
\addplot [darkorange25512714]
table {%
39 8.93333333333334
41 8.93333333333334
};
\addplot [darkorange25512714]
table {%
47 9.21666666666667
49 9.21666666666667
};
\end{axis}

\end{tikzpicture}

        \caption{Latency for \gls{vr} user with different allocations}
        \label{fig:vr_ue_2}
    \end{subfigure}
	\begin{subfigure}[b]{.5\linewidth}
	    \centering
\begin{tikzpicture}[scale=0.85]

\definecolor{darkgray176}{RGB}{176,176,176}
\definecolor{darkorange25512714}{RGB}{255,127,14}

\begin{axis}[
width=\tfwidth,
height=\tfheight,
legend cell align={left},
legend style={fill opacity=0.6,font={\tiny} ,draw opacity=1, text opacity=1, draw=lightgray204},
tick align=outside,
tick pos=left,
x grid style={darkgray176},
xlabel={Allocation type},
xmin=-2, xmax=52,
xtick style={color=black},
xtick={0,8,16,24,32,40,48},
xticklabels={10RBG, 12RBG, 15RBG, 17RBG, 18RBG, 20RBG, 10ms},
y grid style={darkgray176},
ylabel={Bitrate (Mbit/s)} ,
ymin=5, ymax=19,
xmajorgrids,
ymajorgrids,
ytick style={color=black},
label style={font=\small},
tick label style={font=\footnotesize}
]

\addplot [black]
table {%
-1 17.750952
1 17.750952
1 18.276568
-1 18.276568
-1 17.750952
};
\addplot [black]
table {%
0 17.750952
0 17.389168
};
\addplot [black]
table {%
0 18.276568
0 18.69
};
\addplot [black]
table {%
-0.25 17.389168
0.25 17.389168
};
\addplot [black]
table {%
-0.25 18.69
0.25 18.69
};
\addplot [black]
table {%
7 15.75264
9 15.75264
9 16.364358
7 16.364358
7 15.75264
};
\addplot [black]
table {%
8 15.75264
8 15.326544
};
\addplot [black]
table {%
8 16.364358
8 16.834128
};
\addplot [black]
table {%
7.75 15.326544
8.25 15.326544
};
\addplot [black]
table {%
7.75 16.834128
8.25 16.834128
};
\addplot [black]
table {%
15 12.039088
17 12.039088
17 12.481776
15 12.481776
15 12.039088
};
\addplot [black]
table {%
16 12.039088
16 11.749264
};
\addplot [black]
table {%
16 12.481776
16 12.866896
};
\addplot [black]
table {%
15.75 11.749264
16.25 11.749264
};
\addplot [black]
table {%
15.75 12.866896
16.25 12.866896
};
\addplot [black]
table {%
23 9.850848
25 9.850848
25 10.223808
23 10.223808
23 9.850848
};
\addplot [black]
table {%
24 9.850848
24 9.598608
};
\addplot [black]
table {%
24 10.223808
24 10.490048
};
\addplot [black]
table {%
23.75 9.598608
24.25 9.598608
};
\addplot [black]
table {%
23.75 10.490048
24.25 10.490048
};
\addplot [black]
table {%
31 8.146384
33 8.146384
33 8.478496
31 8.478496
31 8.146384
};
\addplot [black]
table {%
32 8.146384
32 7.88808
};
\addplot [black]
table {%
32 8.478496
32 8.744752
};
\addplot [black]
table {%
31.75 7.88808
32.25 7.88808
};
\addplot [black]
table {%
31.75 8.744752
32.25 8.744752
};
\addplot [black]
table {%
39 5.419616
41 5.419616
41 5.635264
39 5.635264
39 5.419616
};
\addplot [black]
table {%
40 5.419616
40 5.285664
};
\addplot [black]
table {%
40 5.635264
40 5.810944
};
\addplot [black]
table {%
39.75 5.285664
40.25 5.285664
};
\addplot [black]
table {%
39.75 5.810944
40.25 5.810944
};
\addplot [black]
table {%
47 7.11104
49 7.11104
49 10.816458
47 10.816458
47 7.11104
};
\addplot [black]
table {%
48 7.11104
48 6.82752
};
\addplot [black]
table {%
48 10.816458
48 17.117856
};
\addplot [black]
table {%
48.75 6.82752
48.25 6.82752
};
\addplot [black]
table {%
48.75 17.117856
48.25 17.117856
};
\addplot [darkorange25512714]
table {%
-1 17.990128
1 17.990128
};
\addplot [darkorange25512714]
table {%
7 16.052928
9 16.052928
};
\addplot [darkorange25512714]
table {%
15 12.244432
17 12.244432
};
\addplot [darkorange25512714]
table {%
23 10.03096
25 10.03096
};
\addplot [darkorange25512714]
table {%
31 8.308688
33 8.308688
};
\addplot [darkorange25512714]
table {%
39 5.523824
41 5.523824
};
\addplot [darkorange25512714]
table {%
47 7.795888
49 7.795888
};
\end{axis}

\end{tikzpicture}
        \caption{Bit rate available to iperf3 user with different allocations}
        \label{fig:data_ue_2}
    \end{subfigure}
    \vspace{-0.5em}
     \vspace{-1em}
     \caption{Latency and bit rate box plots of static allocations (10RBGs - 20RBGs) as well as the data-driven allocation (10ms).}
     \vspace{-1.5em}
 \label{fig:test2}
\end{figure*}

Next, we apply slice isolation to create a set of dedicated resources that are allocated to the \gls{vr} user. The iperf3 user gets the remaining resources (with a total of 25 \glspl{rbg} to share). Before presenting the remaining results in Fig. \ref{fig:data_ue_1}, we consider static slice allocation, where a set of \glspl{rbg} are allocated to a slice, and a \gls{ue} is associated with the slice. In this context, we create one custom slice, which is intended to assure the \gls{vr} user a specific performance. This is tested with six allocations to study the impact. The \gls{vr} \gls{ue} is allocated 10, 12, 15, 17, 18, and 20 \glspl{rbg} respectively in each experiment. The other \gls{ue} is on a best-effort slice, that simply utilizes the remaining \glspl{rbg} available. 
In Fig. \ref{fig:vr_ue_2} a box plot of the performance shows the impact in the respective experiments, with the whiskers showing the 5th and 95th percentiles. As expected, the more \glspl{rbg} allocated to the \gls{vr} user the lower the experienced latency is. On the other hand, as shown in Fig. \ref{fig:data_ue_2}, the fewer \glspl{rbg} allocated to the best effort slice, the less bit rate can be consumed by the iperf3 user.
This means that, while an improved latency is experienced on average, the large variation in frame size means that too many resources are dedicated for time instances with small frame sizes, hence those resources remain unused.

To prevent this wastage of resources, we apply the data-driven approach to monitor the \gls{vr} traffic and dynamically update the allocation. The desired latency that the xApp strives to reach is requested to be ten milliseconds, with the one-millisecond slack.
From Fig. \ref{fig:vr_ue_3}, the data-driven approach is shown. While the latency increases above 10 ms briefly around the 280-second mark the latency quickly returns to oscillating just below the 10 ms target. 
Interesting to the other \gls{ue}, is how bandwidth is allocated when the \gls{vr} traffic does not need it anymore. For example from 150 seconds until 280 seconds, an increase in throughput is made available to the iperf3 user. As the \gls{vr} application is observed to require a smaller data rate to maintain the latency, additional \glspl{rbg} are utilized for the iperf3 user instead.
For the data-driven slicing the mean experienced VR latency was 9.3 milliseconds. To achieve the same latency using static allocation would require 18 \glspl{rbg}.
We found that the mean bit rate of the iperf3 user is 8.2 and 9.5 Mbit/s when applying the static allocation of 18 \glspl{rbg} and when applying data-driven slicing, respectively. This means more resources were made available to the iperf3 user while maintaining the same latency for the VR user, demonstrating how data-driven slicing provides a more efficient use of resources. While the data-driven approach can improve secondary user data rate while guaranteeing latency of the VR user, it is worth noting that this guarantee costs $\sim 17$ Mbit/s, when comparing the total achieved bit rate of $\sim 34$ Mbit/s and $\sim 17$ Mbit/s in Fig. \ref{fig:vr_ue_3} (b) and (d), respectively. However, for a system with more VR users allocated in the same slice, the slicing overhead would be less.

\section{Conclusion and Outlook}
In this work we have experimentally demonstrated data-driven \gls{ran} slicing, which uses a continual adaptation of \gls{rb} allocation to ensure low latency for a \gls{vr} stream in a resource-constrained two-user scenario. This is realized in an ORAN-based testbed, allowing monitoring of the \gls{ran}, where we extract radio-level information for specific users. Data-driven slicing is implemented as an xApp that creates a \gls{ran} slice for a \gls{vr} user, and instantiates monitoring and control of the performance of this \gls{ue}. 
From the monitoring data, the xApp uses a heuristic algorithm to identify the individual video frames and the duration of their transmission from which the RAN latency is estimated. The average estimated latency over a 60 frame period is used to decide if more or fewer \glspl{rbg} should be allocated to reach the target of 10 ms latency.
Compared to the static \gls{rbg} allocation that would result in a similar VR latency, the data-driven allocation allows the secondary user to have a $15.8\%$ higher throughput. 

An obvious next step is to replace the heuristic rule-set of the data-driven \gls{rb} allocation algorithm by an AI algorithm, e.g., exploiting the reinforcement learning paradigm to enable the xApp to adapt to the characteristics of different types of application traffic without manually specifying a corresponding ruleset.
Another direction could be to optimize the perceived \gls{qoe} instead of VR latency, to avoid unnecessary overprovisioning. As QoE cannot be measured directly, proxy metrics would need to be identified and, e.g., learned by a neural network, which could in turn be used to guide resource allocation.

\bibliographystyle{IEEEtran}
\bibliography{references}

\end{document}